\documentstyle[epsf,12pt]{article}
\begin{document}
\def\salto{\par\vskip .5cm}
\date{}
\author{C. BECCHI}
\title{INTRODUCTION TO GAUGE THEORIES}
\maketitle

\salto
\centerline{\it Dipartimento di Fisica, Universit\`a di Genova,}
\centerline{\it Istituto Nazionale di Fisica Nucleare, Sezione di Genova,}
\centerline{\it  via Dodecaneso 33, 16146 Genova (Italy)}
\salto
\noindent
\abstract{These lectures present an elementary introduction to quantum
gauge fields.
The first aim is to show how, in the tree approximation, gauge invariance
follows from covariance and unitarity.  This leads to the standard
construction of the Lagrangian by means of covariant derivatives in a form
that unifies the massive and the massless case.
Having so identified the classical theory, the Faddeev-Popov quantization
method is introduced and the BRS invariance of the resulting action is
discussed. }
 \vfill \footnote{Lectures given at
the Triangle Graduate School 96, Charles University, Prague  September 2-11,
1996}

\eject

\def\salto{\par\vskip .5cm}
\def\.{\cdot}
\def\la{\lambda}
\def\s{\sigma}
\def\bs{{\bar\sigma}}
\def\t{\tau}
\def\o{\over}
\def\v{\vec}
\def\a{\alpha}
\def\z{\zeta}
\def\c{\gamma}
\def\b{\beta}
\def\d{\delta}
\def\k{\chi}
\def\f{\phi}
\def\vf{\varphi}
\def\C{\Gamma}
\def\S{{\bf S}}
\def\P{\Psi}
\def\T{\Theta}
\def\La{\Lambda}
\def\O{\Omega}
\def\x{\xi}
\def\n{\eta}
\def\u{\omega}
\def\ub{{\bar \u}}
\def\o{\over}
\def\p{\partial}
\def\ip{\int {d p\o  (2\pi)^4}}
\def\inx{\int d^4 x}
\def\iny{\int d^4 y}
\def\inz{\int d^4 z}
\def\dv{d \vec}
\def\ne{\not=}
\def\+{\bigoplus}
\def\D{\Delta}
\def\fo{{\cal F}_0}
\def\oo{{\cal O}}
\def\fc{{\cal F}_C}
\def\({\left(}
\def\){\right)}
\def\[{\left[}
\def\]{\right]}
\def\l.{\left.}
\def\r.{\right.}
\def\sec{\section}
\def\ss{\subsection}
\def\be{\begin{equation}}
\def\ee{\end{equation}}
\def\bea{\begin{eqnarray}&&}
\def\eea{\end{eqnarray}}
\def\nn{\nonumber \\ &&}
\def\nnn{\nonumber \\ }
\def\acca{\right.\nn\left.}
\def\ber{\begin{array}}
\def\eer{\end{array}}
\def\ca{{\vec {\cal A}}}

\renewcommand{\theequation}{\arabic{equation}} \newcommand{\e}{\mbox{e}}
\newcommand{\g}{\mbox{g}} \renewcommand{\arraystretch}{2.5}
\newcommand{\bra}{\langle}
\newcommand{\ket}{\rangle}
\def\eins{  1\!{\rm l}  }
\salto
\sec{Introduction}
\salto
The recent developments in high energy physics have put great emphasis
 on gauge theories; indeed the general theory of fundamental interactions
 is completely formulated in this framework. The importance of the role of
 gauge invariance has obscured the reasons that have historically justified the
 introduction and the development of gauge theories as consistent field
 theories. Indeed they are very often justified on the basis of a
 "symmetry principle" that has to be accepted as a fundamental principle
 in nature.

As a matter of fact great collective efforts have been needed to identify
 gauge theories as the natural, and in a sense unique, quantum theories
 of vector fields. It is the primary role of vector currents hiding
associate vector fields and the ensuing discovery of the corresponding
 bosons that has put gauge theories in their actual preeminent position
 in High Energy Physics. The progresses in the understanding of the
fundamental nature of strong interactions is also based on gauge theories
in spite of the lack of associate observable charges and vector particles.
 The hypothesis of confinement has in fact extended the range of gauge
theories  opening the possibility that the vector fields
 give an adequate  description only of the short
distance properties of strong interactions \cite{bu}. They should  remaining
unobservable together with the associated charges since
 at large times and distances  their presence is hidden  by a catastrophic
growth of the strength of the coupling,.
 In spite of
the lack of vector field asymptotic states, gauge invariance appears to
 play an essential role in the control of the strength of interactions
 at short distances. .
The scale properties of scattering amplitudes
 at large momenta giving strong indications of a weakening of the coupling
strength single out a gauge theory (QCD) among the possible field theoretical
models.

On the basis of these considerations I have tried, in the first part of these
lectures, to give an introduction  to vector
field theories based on the fulfillment of physical consistency criteria
 among which very important is the unitarity of scattering amplitudes.
The main lines of this analysis are the same followed by Yang and Mills
in their fundamental paper  \cite{ym}. I am just presenting a translation
of  the
same analysis into the modern functional language.

The second main part of the lectures are devoted to the construction of gauge
theories as fundamental local field theories. This will be done  discussing the
structure of the functional measure involved into the Feynman functional
integral form of the vacuum-to-vacuum transition amplitude.

As a matter of fact there are two complementary measures that are commonly
taken into account in the construction of gauge theories. The first one is
the Wilson measure \cite{2} that is based on the lattice regularization. This is
a  regularization method of field theories in which the space-time points are
 identified with those of a periodic four-dimensional lattice.
The number of these points being finite, the functional integral is reduced
 to an ordinary integral. The Wilson measure does not involve vector fields; it
is suitable for the description of the theory in the strong coupling
regime, that is, in the case of QCD, at large distances. We shall not discuss
Wilson measure but those belonging to the second class, the vector field
theories. These are the natural generalisation of quantum electrodynamics and
 the basis of all the perturbative calculations; in QCD they are suitable
for a description of the theory at short distances and  responsible
for the introduction of important concepts, as e.g. that of "gluon".

The lectures are organized as follows: in the first lecture the main tools and
results of field theory in its functional formulation are reminded. In the
second lecture vector field theories are introduced presenting the arguments
leading to gauge invariance. In the third lecture the construction of vector
effective field theories is described. In the fourth lecture we discuss the
Higgs mechanism giving origin to massive vector particles. The fifth lecture is
devoted to the construction of gauge theories in  Feynman functional framework.
In the last lecture we describe the symmetry properties of this theory.

 \salto
\sec{The S-matrix}
\salto
In quantum field theory \cite{1} the scattering amplitudes are
computed by means of the reduction formula. This can be simply written using
the Green functional  generator of the theory that is defined according:
\be Z[j]\equiv e^{{i\o\hbar}Z_c[j]}=
<\O,T\( e^{{i\o\hbar}\inx \sum_\a\f_\a (x) j_\a(x)}\) \O>\
 .\label{1}\ee
where $\f_\a$ and $j_\a$  label a set of quantized fields  and
corresponding sources with, in general, different Lorentz covariances. $Z_c$ is
the connected functional. Whenever one is interested in computing matrix
elements between scattering states of composite operators, that is of operators
that are  given by non-linear functions of fields, this is done by introducing
a "source" $\z_i$ for each operator $O_i$ and extending the definition of $Z$,
and hence of  $Z_c$, according:
\be  Z[j,\z]=<\O,T\( e^{{i\o\hbar}\inx\[ \sum_\a\f_\a (x) j_\a(x)+
\sum_i\z_i(x)O_i(x)\]}\) \O>\
 .\label{1a}\ee
The "connected" n-point functions are given by the n-th functional derivatives
of $Z_c$; in particular the  two-point function is:
\be{\d^2\o\d j_\a(x)\d j_\b(0)}Z_c|_{j=0}\equiv \D^{\a\b}(x)\ .\label{3}\ee
In this summary of the main tools and results of the quantum field theory of
scattering
 we shall limit ourselves for simplicity to the case of massive fields;
indeed with massless fields one has long range forces that change drastically
the nature of the scattering states. In the massive case  the asymptotic
particles correspond to poles of the Fourier transformed Green functions. In
particular  we can separate from $\D$ the asymptotic propagator $\D_{as}$: \be
\D^{\a\b}(x)=\sum_\la\ip{e^{{i\o\hbar}px}\o m_\la^2-p^2-i0_+}\C_\la^{\a\b}(p)
+R(x)\equiv \D^{\a\b}_{as}(x)+R(x)\ ,\label{4}\ee where the Fourier transform
of $R$ has no pole in $p^2$. It is clear that the asymptotic propagator is by
no means unique since $\C_\la^{\a\b}$ is defined up to a polynomial in $p^2$
vanishing at $m_\la^2$; however this lack of uniqueness does not affect the
$\S$ matrix that is obtained through the LSZ reduction formulae.

Given
$\C_\la^{\a\b}(p)$ one
 introduces the asymptotic free fields
$\f_{in}$ with the commutation relations:
\be\[\f_{in}^{\a (+)} (x),\f_{in}^{\b (-)}
(0)\]=\sum_\la\ip e^{{i\o\hbar}px} \theta (p^0)\d (p^2-m_\la^2)\C_\la^{\a\b}(p)
\ ,\label{5}\ee
and the asymptotic wave operator: \be K_{\a\c}(\p)\D^{\c\b}_{as}(x)=
\d_\a^\b\d (x)\
,\label{6}\ee  the $\S$ matrix is given in the asymptotic Fock space by:
\be \S =:e^{{1\o\hbar}\inx\f^\a (x) K_{\a\b}(\p ){\d\o\d j_\b (x)}}:
Z  |_{j=0}\equiv
:e^\Sigma : Z
|_{j=0}\
.\label{7}\ee

 Thus the whole  dynamical information is contained into the Green functional
$Z$. This is computed by means of the Feynman formula: \be Z[j,\z]=\int d\mu\
e^{{i\o\hbar}\inx \[\f (x) j(x)+ \sum_i\z_i(x)O_i(x)\]}\ ,\label{Fi}\ee in terms
of the functional measure of the theory $d\mu$  that is related to the "bare"
action $S$ by the heuristic relation;
\be d\mu=N\prod_x d\f (x) e^{i {S (\f )\o\hbar}}\ .\label{er}\ee
 $N$ is a normalization factor implementing the normalization condition:
$Z[0,0]=1$.

The heuristic formula (\ref{Fi}) acquires a well definite meaning after a
regularization-renormalization procedure which is systematically known only at
the perturbative level. The standard procedure in perturbative QCD is based on
dimensional regularization. Due to obvious time limitations, we shall disregard
this, however important, step of the construction.

 It is reminded above that the functionals $Z$ and its
connected part $Z_c$  are directly related to the scattering amplitudes; the
quantum analog of  the action is given by the Legendre transform of $Z_c$,  the
proper functional $\C$ \cite{be}. Perturbatively this  is the functional
generator of the 1-particle-irreducible  amplitudes, that is of   the amplitudes
corresponding to Feynman diagrams that cannot be divided into two disconnected
parts by cutting a single line, it is often called the effective action,
although this name is also  shared by completely
different objects.

 To introduce the proper functional one defines the field functional:
\be\f \[j,\z,x\]\equiv{\d\o\d j(x)}Z_c\[j,
\z\]- {\d\o\d j(0)}Z_c\[0,0\]\ ,\label{10}\ee
then, assuming that the inverse functional $j\[\f,\z ,x\]$
be uniquely defined, one has:
\be\Gamma\[\f,\z\]\equiv Z_c\[j\[\f,\z\],\z\]-
\int dx\(\f (x)+{\d\o\d j(0)}Z_c\[0,0\]\) j\[\f,\z ,x\]\ .\label{11}\ee
It is easy to verify that:
\be{\d\o\d \f (x)}\Gamma\[\f \[j,\z\],\z\]=-j\[\f,\z ,x\]\
,\label{pm}\ee and
\be {\d\o\d \z (x)}\Gamma\[\f ,\z\]|_{\f=
\f \[j,\z\]}={\d\o\d \z (x)}Z_c\[j,\z\]\ .\label{est}\ee
Therefore:
\be {\d^2\o\d \f\d\f'}\Gamma\[\f ,\z\]|_{\f=
\f \[j,\z\]}=-\[{\d^2\o\d j\d j'}Z_c\[j,{\z}\]\]^{-1}\ . \label{pro}\ee
That is: the second field-derivative of $\Gamma$ gives the "full" wave operator
(not to be mistaken with the asymptotic one defined in (\ref{6})).
Notice that the $-$ sign in (\ref{pm}) refers to bosonic fields while in the
fermionic case one has the opposite sign.

Perturbation theory consists in the construction of $\C$, $Z_c$ and $S$ as
(formal) power series in $\hbar$.  The fundamental tool of this
construction is the saddle point method. Considering the leading
steepest descent contribution to the integral (\ref{Fi}) we get: \be
Z_c=[S+j\f+\z O ]|_{{\d S\o\d \f}+j+\z{\d O\o\d \f}=0 , \hbar=0}+O(\hbar ) \
,\label{thr}\ee and hence, if the equation ${\d S\o\d \f}|_{\hbar=0}=0$ has the
unique solution: $j=0$, we have:
\be \C=[S+\z O]|_{\hbar=0}+O(\hbar )\ .\label{16}\ee
Therefore in the "classical limit" $\C$ coincides with the action extended
to take into account the source terms of the composite operators.

This concludes our general introduction to the functional approach to field
theory.
In the following sections we shall apply the above results to vector
field theories.

\sec{Vector fields and gauge invariance}
\salto
In the study of vector field theories one has to
distinguish the case of massive from that of
massless fields and theories in which the vector fields are self-coupled
(non-abelian) from those without self couplings (abelian).
In particular, abelian theories behave at
large times as free ones, therefore describing scattering processes
 of spin 1 particles. The massive case differs from the massless one for
the asymptotic state counting since massless vector particles have two
 helicity states while the massive one have three of them.

The non-abelian situation is much more involved; indeed the massive case
 originates from the so-called Higgs mechanism, corresponding to an highly
non-trivial structure of the vacuum state. The vector particles are analogous to
quanta of the plasma oscillations in a medium containing free charged particles;
hence the vacuum is analogous to a condensate of "charged particles".
In much the same way as in the abelian case, in the Higgs theory the coupling
weakens at  large distances and hence one has a scattering theory involving
asymptotic
 spin 1 states.

In a non-abelian massless vector field theory one faces a completely
different situation. Indeed, if the  number of matter fields is small,
 the coupling tends to weaken at short distances,
while it is supposed to intensify at large distances. This phenomenon, that
would
lead to the confinement of the charged states, and hence no scattering
theory for
them, is interpreted from the fundamental point of view as a condensation of
"magnetic monopoles" rather than charges. This would be a mechanism dual to
superconductivity \cite{ds}, confining the field strength into thin flux tubes
and hence giving a constant attractive force between two opposite charged
particles. Contrary to the massive case, in this situation vector fields would
play a role only at short distances where they are associate with a very
peculiar force law.  For our purposes it is convenient to limit our analysis,
at least at the beginning, to the massive case in which it is possible
to exploit the whole apparatus described in the previous section.

Let us therefore consider a system of vector fields $ A^\a_{\mu}$
that are assumed to have non-trivial asymptotic limit
 leading to scattering. Let the corresponding asymptotic fields be:
\be A^{(in)\a}{\mu}(x)= \frac{1}{(2\pi)^{3/2}}\sum_{\lambda=1}^3
\int\frac{d^3p}{\sqrt{2E_p}}\left( \epsilon_{\lambda,\vec{p},\mu}
{\cal A}^{(in)\a}_{\lambda\vec{p}} e^{-{ipx\o\hbar}}+hermitian \
\  conjugate\right) \ ,\label{asy}\ee
with the polarization vectors $\epsilon$ satisfying:
\be  \epsilon_{\lambda,\vec{p},\mu} \epsilon^{\mu}_{\lambda ',\vec{p}}
=-\d_{\lambda\lambda '}\ ,\label{pol}\ee and
\be  \epsilon_{\lambda,\vec{p},\mu} p^{\mu}=0\ .\label{tr}\ee
Notice that:
\be \partial_{\mu}A^{(in)\a \mu}(x)= -\frac{i}{(2\pi)^{3/2}}\sum_{\lambda=1}^3
\int\frac{d^3p}{\sqrt{2E_p}}\left( \epsilon^{\mu}_{\lambda,\vec{p}}
p_{\mu}{\cal A}^{(in)\a}_{\lambda,\vec p}e^{-{ipx\o\hbar}} - h. c. \right)=0
\ .\label{tran}\ee
Here and in the following to simplify the formulae, we assume mass
degeneracy of the vector field components.

From (\ref{asy}) the asymptoptic propagator is easily computed:
\be \D_{as\mu\nu}^{\a\b}(x)=
\ip{e^{{i\o\hbar}px}\[{p_\mu p_\nu\o m^2}-g_{\mu\nu}\]
\o m^2-p^2-i0_+}\d^{\a\b}\ .\label{ap}\ee
The two-point function is:
\be \D_{\mu\nu}(x)=\D_{as\mu\nu}(x) + R_{\mu\nu}(x)\ ,\label{2p}\ee
where $R$ accounts for the contributions of many particle states in the Lehmann
decomposition of $\D$.

The main difficulty with this theory comes from the high momentum behavior
of the Fourier transform of the asymptotic propagator that tends to an
homogeneous
function of degree zero:
\be \tilde \D_{as\mu\nu}^{\a\b}(p)\longrightarrow -{p_\mu p_\nu\o p^2 m^2}
\d^{\a\b}\ ,\label{limprop}\ee
instead of $-2$ that is expected for dimensional reason. The same behavior
 characterizes the two-point function $\D$ at least not too far from
the perturbative regime. It is fairly obvious that the propagator of degree zero
is going to produce cross sections increasing proportionally to the second
power
of the center-of-mass energy \cite{ls}
in much the same way as this happens in
the Fermi theory of weak interactions. This induces a violation of $\S$-matrix
unitarity; therefore, in order the theory to be consistent, one has  to assume
a decoupling mechanism for the  "longitudinal" components of the vector field.
We are going to show that this leads directly to gauge invariance.

A convenient, however elaborate,
way to analyze this decoupling mechanism is based on the introduction
of artificial longitudinal asymptotic states by adding in
 (\ref{asy}) a further set
 of polarization states corresponding to $\la=0$ with:
\be  \epsilon_{0,\vec{p},\mu}\sim p^{\mu}\ .\label{lon}\ee
A direct way to do this is to add to the proper functional $\C$ the term:
\be -\inx{\xi\o 2}\sum_\a\[\partial_{\mu}A^{\a \mu}(x)\]^2\ .\label{xi}\ee
The Fourier transformed two-point vertex before the introduction of
(\ref{xi}) is:
\be\tilde\C^{(2)\a\b}_{\mu\nu}(p)=\d^{\a\b}\[\[{p_\mu p_\nu\o p^2}-g_{\mu\nu}\]
\(p^2-m^2\)\(1+A(p^2)\)
+{p_\mu p_\nu m^2\o p^2}\(1+B(p^2)\)\]\ ,\label{aza}\ee
with $A(0)=B(0)$ since (\ref{aza}) must be regular at $p^2=0$ and,
owing to (\ref{ap}), $A(m^2)=B(m^2)=0$. From (\ref{aza}) we can compute the
Fourier  transformed two-point function after introduction of
(\ref{xi}) getting:
\be\tilde\D^{(2)\a\b}_{\mu\nu}(p)=\d^{\a\b}\[{\[{p_\mu p_\nu\o p^2}-g_{\mu\nu}\]
\o
\(m^2-p^2\)\(1+A(p^2)\)}
+{p_\mu p_\nu\o p^2\(\xi p^2-m^2\(1+B(p^2)\)\)}\]\ .\label{prop}\ee
(\ref{prop}) has a pole at $p^2=m^2$ in much the same way as the Fourier
transform of (\ref{ap}), however its degree at high momentum is $-2$ and
 there is a second pole at $p^2={m^2\o\xi}\(1+B(p^2)\)$.
 Considering the asymptotic
 propagator corresponding to this second pole, we see that this can be
identified with
 that of the derivative of a scalar field, however with the wrong sign.
This shows that the introduction into the vertex functional of the term
(\ref{xi}) is just a mathematical trick void of any physical meaning whose
only role is to "regularize" the short distance properties of the two-point
function. Indeed the wrong sign indicates that the modified proper functional
does not correspond to any quantum field theory whose asymptotic state space
be a Hilbert space; the sign is however consistent with a theory in an
indefinite metric space.

Our purpose is to prove that the fake theory with
the term (\ref{xi}) leads to an $\S$ matrix independent
of $\xi$, which  is equivalent to the wanted decoupling of
the  "longitudinal" components
of the vector fields.  Since ,
this theory is expected to give cross sections less divergent at high energies,
 and hence compatible with unitarity, and since, in the limit
$\xi\rightarrow 0$, one gets back the original scattering amplitudes, the
$\xi$-independence of the  $\S$ matrix of the  fake theory excludes the feared
violations of unitarity in the original one.

The decoupling of
the  "longitudinal" components
of the vector fields is equivalent to the equation:
\be : \partial_{\mu}A^{\a \mu}(x)
|_{\f={\d Z_c\o\d j} , j=K\f_{in}}:=\p_\mu A^{\a \mu}(x)_{in}\ .\label{i4}\ee
Now, using the field equation (\ref{pm}) on the mass-shell, we have:
\be : \[\xi \p_{\mu}\partial_{\nu}A^{\a \nu}(x)+{\d\C\o\d A^{\a \mu}(x)}\]
|_{\f={\d Z_c\o\d j} , j=K\f_{in}}:=0\ ,\label{i5}\ee
whence we see that (\ref{i4}) is equivalent to:
\be : \p^{\mu}{\d\C\o\d A^{\a \mu}(x)}
|_{\f={\d Z_c\o\d j} , j=K\f_{in}}:=0\ ,\label{ii}\ee
in the subspace of the asymptotic space in which $\p_\mu A^{\a \mu}(x)_{in}=0$.

In the forthcoming formulae
 we shall forget the Wick ordering and understand the mass shell
prescription appearing  above  whenever we shall write an identity
using the identity symbol $=$. We shall instead use the equivalence
symbol
$\equiv$ in the case of  relations holding true even outside the mass
shell.

(\ref{ii}) shows that ${\d\C\o\d A^{\a \mu}(x)}$
is a conserved current. After N\"oether theorem, given a
system of conserved currents $I^{\mu\a}$,
one can always find a set of infinitesimal field transformations:
\be
\f^i(x)\Rightarrow \f^i(x) + \epsilon_{\alpha}
 P^i_\a(x)\ \label{tra}\ee
leaving $\C$ invariant and hence such that:
\be P^i_\a(x) \frac{\delta \Gamma}{\delta \f^{i}} \equiv
\partial_\mu I^{\mu \alpha}\equiv
\p^{\mu}{\d\C\o\d A^{\a \mu}}\ .\label{inv}\ee
In (\ref{tra})  the symbols $P^i_\a$
 stay for generic field functionals
and once again we have used $\f$ as a collective symbol for all the fields.

Adding to $\C$ the $\xi$ term and applying to the result the transformations
(\ref{tra}) we get on the mass shell:
\be -\xi\p_{\mu}A^{\b\mu}\p_{\nu}P^{\b\nu}_\a+ P^i_\a
\frac{\delta \Gamma}{\delta \f^{i}}=0\ ,\label{i6}\ee
and hence, on account of (\ref{i5}) and (\ref{inv}) we find:
\be \xi\[\p^2\d^{\b}_\a+\p^{\nu}P^{\b\nu}_\a\]\p^{\mu}A_{\b\mu}=0\
 ,\label{fp}\ee
in the  asymptotic subspace in which $\p_\mu A^{\a \mu}(x)_{in}=0$.

Therefore, provided that the kernel of the differential operator
$\[\p^2\d^{\b}_\a+\p^{\nu}P^{\b\nu}_\a\]$ be trivial, we can conclude that
(\ref{inv}) implies (\ref{i4}) and hence the $\xi$-independence of the
$\S$ matrix. Notice that we can write (\ref{inv}) in the form:
\be \left( \partial_\mu \frac{\delta }{\delta A_\mu^{\alpha}} - P_{\alpha}^{i}
\frac{\delta }{\delta \phi_{i}}\right) \Gamma \equiv X_{\alpha}(x) \; \Gamma
\equiv 0\ ,  \label{gi} \ee

It is clear that  (\ref{gi}) corresponds to an invariance property of $\C$.
Considering in particular the case in which the  fields $\f$ correspond to
the vectors and to a system of scalars $\vf^a$, that is:
\be \f^{i} \equiv \left( A^{\beta\mu},\vf^{a} \right) \ ,\label{gt1}\ee and:
\be P_\alpha^{i} \equiv \left(
P_{\alpha }^{\beta\mu},P_{\alpha}^{a}\right) \ ,\label{gt2} \ee
(\ref{gi}) prescribes the invariance of $\C$ under the system of
 infinitesimal transformations:
\be A^{\alpha\mu}(x) \rightarrow A^{\alpha\mu}(x) +
\partial^\mu \Lambda^{\alpha}(x)
+ \Lambda^{\beta}(x)P_{\beta}^{\alpha \mu}(x) \ ,\label{gt3}\ee
and
\be \vf^{a}(x) \rightarrow \vf^{a}(x) + \Lambda^{\beta}(x)
P_{\beta }^{a}(x) \ .\label{gt4}\ee
We shall call these transformations "gauge transformations" and
the invariance condition (\ref{inv}) "gauge invariance condition".

To push farther our analysis we assume that the above described invariance
be minimal, in the sense that no further condition is needed for the theory
 to be consistent. The meaning of this hypothesis is clarified in mathematical
terms requiring that the system of differential operators $X_{\alpha}(x) $ be in
involution, that is:
\be \left[ X_{\alpha}(x), X_{\beta}(y) \right]
=\int dz F_{\alpha\beta}^{\gamma}(x,y;z)
 X_{\gamma}(z)\ . \label{li}\ee
Indeed, according to Frobenius theorem, the commutation relations
 (\ref{li}) are necessary and sufficient conditions for the system (\ref{gi})
be integrable.

Our aim is now to use (\ref{li}) to get further information on the nature
of the transformations (\ref{gt3}) and (\ref{gt4}). In the framework of an
effective field theory one studies the low energy properties of a quantum
system;
if all the particles are massive, in the low energy regime,
 the vertices of the theory are analytical functions of the momenta;
this implies that they can be approximated by polynomials and hence
$\C$ is approximately a local functional. The first non-trivial approximation
contains only the terms whose coefficients have non-negative mass dimension;
further terms with coefficient of increasingly high negative mass dimension are
needed if one studies processes of increasing energy.
The same considerations hold
true for $P^i_\a$ whose first approximation,
on account of Lorentz invariance, is:
\be P^{\beta \mu}_{\alpha} (x) = C^{\beta}_{\alpha \gamma}
A^{\gamma\mu}(x)\ , \label{gt5}\ee
\be P_{\alpha}^{a} (x) = v_{\alpha}^{a}+t_{\alpha}^{ab} \vf^{b}(x)
\ . \label{gt6} \ee
Let us consider, for the moment the case in which the gauge
transformations act homogeneously on the scalar fields, that is:
\be  v_{\alpha}^{a}=0\ , \label{gt7}\ee and hence:
\be X_{\alpha} (x)= \partial_{\mu} \frac{\delta}{\delta A_{\mu}^{\alpha}} -
C_{\alpha \gamma}^{\beta} A_{\mu}^{\gamma}(x)
\frac{\delta}{\delta A_{\mu}^{\beta}} - t_{\alpha}^{a b} \vf^{b}
\frac{\delta}{\delta \vf^{a}}\ .\label{gt8}\ee
In this situation (\ref{li}) is written:
\begin{eqnarray}
\left[ X_{\alpha}(x), X_{\beta}(y) \right] & = &
\underbrace{-\left[ \partial_{\mu} \frac{\delta}{\delta A_{\mu}^{\alpha}(x)},
C_{\beta \delta}^{\gamma} A_{\nu}^{\delta}(y)
\frac{\delta}{\delta A_{\nu}^{\gamma}(y)} \right]}_{1} +
\underbrace{\left( \begin{array}{ccc}
x & \Leftrightarrow & y \\
\alpha & \Leftrightarrow & \beta \end{array} \right)}_{2} + \nonumber \\
 &  & + \underbrace{\left[ C_{\alpha \delta}^{\gamma} A_{\mu}^{\delta}(x)
\frac{\delta}{\delta A_{\mu}^{\gamma}(x)} ,
C_{\beta \zeta}^{\eta} A_{\nu}^{\zeta}(y)
\frac{\delta}{\delta A_{\nu}^{\eta}(y)} \right]}_{3} + \nonumber \\
 &  & + \underbrace{\left[ t_{\alpha}^{a b} \vf^{b}
\frac{\delta}{\delta \vf^{a}} ,
t_{\beta}^{c d} \vf^{d} \frac{\delta}{\delta \vf^{c}} \right] }_{4}
\ .\label{cc}\end{eqnarray}
The first two terms in the right-hand side give:
\be (1) - (2) =- \partial_{\mu} \delta (x-y) \left( C_{\beta \alpha}^{\gamma} +
C_{\alpha \beta}^{\gamma} \right) \frac{\delta}{\delta A_{\mu}^{\gamma}(y)} +
\delta (x-y) \left( C_{\alpha \beta}^{\gamma} \partial_{\mu}
\frac{\delta}{\delta A_{\mu}^{\gamma}(x)} \right) \ .\label{cc1}
\ee
On account of (\ref{li}) the first term in the right-hand side has to vanish
since in no way it can appear in the right-hand side of  this equation.
Thus we have:
\be C_{\beta \alpha}^{\gamma} + C_{\alpha \beta}^{\gamma} = 0 \ \ee
 taking into account the surviving term in (\ref{cc1}) we see that:
\be  F_{\alpha\beta}^{\gamma}(x,y;z)= \delta (x-z) \delta (y-z)
C_{\alpha\beta }^{\gamma}\ \ee
and hence  (\ref{li}) is written:
\be \left[ X_{\alpha}(x),X_{\beta}(y) \right] = \delta (x-y)
C_{\a\beta }^{\gamma} X_{\gamma}(x)\ . \label{li2}\ee
The third term in (\ref{cc}) is:
\be (3) = \delta (x-y) \left( C_{\alpha \delta}^{\gamma}
C_{\beta \gamma}^{\eta} - C_{\beta \delta}^{\gamma}
C_{\alpha \gamma}^{\eta} \right) A_{\mu}^{\delta}
\frac{\delta}{\delta A_{\mu}^{\eta}} \ ,\ee
and consistency with (\ref{li2}) requires:
\be
C_{\alpha \delta}^{\gamma} C_{\gamma \beta}^{\eta} +
C_{\delta \beta}^{\gamma} C_{\gamma \alpha}^{\eta} +
C_{\beta \alpha}^{\gamma} C_{\gamma \delta}^{\eta} \ .\label{jac}\ee
This is a Jacobi identity; it proves that $ C_{ \beta\gamma}^{\a}$ are
the structure constants of a Lie group the we shall call the "gauge group".
Coming to the fourth term we get:
\be (4) = t_{\alpha}^{ab} t_{\beta}^{ca} \vf^{b}
\frac{\delta}{\delta \vf^{c}} - \left( \alpha \leftrightarrow \beta \right) =
-C_{\alpha\beta }^{\gamma} t_{\gamma}^{cb} \vf^{b}
\frac{\delta}{\delta \vf^{c}} \ee
and from (\ref{li2}):
\be \left[ t_{\a}, t_{\b} \right] =
C_{\alpha\beta }^{\gamma} t_{\gamma} \ .\label{li3}\ee
This identifies the matrices $t$ with the infinitesimal generators of a
representation of the gauge group.

We have thus characterized completely the structure of the gauge
 generators identifying the coefficients $C$ with the structure constants of
the gauge group and the matrices $t$ with a representation of the corresponding
Lie algebra.

If we now relax condition (\ref{gt7}) replacing:
\be X_{\alpha} \rightarrow
X_\alpha - v_{\alpha}^{a} \frac{\delta}{\delta \vf^{a}} \ .\label{hig}\ee
 we insert in the right-hand side of
(\ref{cc}) the further term:
\be \left[ X_{\alpha}(x),X_{\beta}(y) \right] = ... +
\underbrace{\left[ v_{\alpha}^{a} \frac{\delta}{\delta \vf^{a}},
t_{\beta}^{cd} \vf^{d}\frac{\delta}{\delta \vf^{c}} \right] -
\left( \alpha \Leftrightarrow \beta \right)}_{5} \ ,\ee
then, applying (\ref{li2}), we get:
\be t_{\beta} v_{\alpha}-t_{\alpha}v_{\beta}=
C_{\beta \alpha}^{\gamma}v_{\gamma} \ .\label{cc3}\ee
In the standard situation in which the gauge group is semisimple
and the representation $t$ is unitary, one can show by purely algebraic means
that (\ref{cc3}) implies:
\be v_{\alpha}^{a} = t_{\alpha}^{ab} v_{b} \ , \label{cc4}\ee
that is:
\be P_{\alpha}^{a}(x) = t_{\alpha}^{ab} ( \vf^{b} + v^{b} )  \ .\label{cc5}\ee
Making use of scale and dimensional arguments
the same result can be extended to any compact gauge group.
When the matter field representation of the gauge group is unitary
it is convenient to replace the antihermitian matrices $t$ by hermitian ones:
\be t_\a\equiv i\t_\a\ ,\ee
writing (\ref{li3}) according:
\be \left[ \tau_\alpha, \tau_\beta
\right] =- i C_{\alpha \beta}^{\gamma} \tau_\gamma
\equiv i f_{\alpha \beta}^{\gamma} \tau_\gamma \ .\label{uni}\ee
We shall conventionally adopt the normalization condition:
\be Tr\(\t_a\t_b\)=\d_{ab}\ ,\label{norm}\ee
With these symbols the infinitesimal gauge
transformations in the "inhomogeneous"
case are written:
\be
 \left\{
\begin{array}{ccc}
\delta A_{\mu}^{\alpha} & = &  \partial_{\mu} \Lambda^\alpha -
\Lambda^\beta A_\mu^{\gamma} f_{\beta \gamma}^{\alpha} \nonumber \\
\delta \vf^{a} & = & i \Lambda^\alpha \tau_\alpha^{ab} \left( \vf + v
\right)^{b}
\end{array}\label{ino} \right.\ee
The homogeneous form is obtained simply setting $v=0$
\salto
We now show how it is possible to construct the vertex functional of a gauge
invariant effective field theory assuming the gauge transformations
(\ref{ino}). The fundamental tool for this construction is given by the
covariant derivative operator; this is a space-time partial
derivative commuting with the gauge
transformations. Let us consider the gauge variation of
 the usual  space-time partial derivative of the scalar field; this is given by:
\be \delta \partial_\mu \phi = i \Lambda^\alpha \tau_\alpha
\partial_\mu \phi + i \partial_\mu \Lambda^\alpha \tau_\alpha \phi \ ,\ee
from which it is fairly clear that this derivative is not covariant.
However consider:
\begin{eqnarray} \delta \left( \partial_\mu \vf
-i A_\mu^{\alpha} \tau_\alpha \( \vf +v\)
\right)& = & i \Lambda^\alpha \tau_\alpha \partial_\mu \vf +
i \partial_\mu \Lambda^\alpha \tau_\alpha\( \vf +v\) - i \partial_\mu
\Lambda^{\alpha} \tau_\alpha \( \vf +v\) \nonumber \\
& & + i \Lambda^\beta A_\mu^{\gamma}
f_{\beta \gamma}^{\alpha} \tau_\alpha \( \vf +v\) + A_\mu^{\alpha}
\tau_\alpha \Lambda^\beta \tau_{\beta} \( \vf +v\) \nonumber \\
 & = & i \Lambda^\alpha \tau_\alpha \left( \partial_\mu \vf -
i A_\mu^{\beta} \tau_\beta \( \vf +v\) \right)
\ .\end{eqnarray}
This shows that the derivative:
\be D_{\mu}\( \vf +v\)\equiv  \partial_\mu \vf
-i A_\mu^{\alpha} \tau_\alpha \( \vf +v\)\ ,\label{cd}\ee
is covariant, since $ D_\mu \( \vf +v\)$ transforms as $\vf+v$.
The same property holds true for a multiple $D$ derivative.

Since the action of a gauge transformation on the scalar fields is unitary,
and hence there is a scalar product $(.,.)$ such that
\be \(\vf +v,\d \( \vf +v\)\) + \(\d \( \vf +v\),\vf+v\)=0\ ,\ee
the scalar products:
\be \(D_{\mu_1}..D_{\mu_n}\(\vf+v\),D_{\nu_1}..D_{\nu_m} \( \vf +v\)\)\ ,\ee
are gauge invariant.

Therefore the functional:
\be\C_s\equiv
\inx\[ \(D_{\mu}\(\vf+v\),D^{\mu} \( \vf +v\)\)-
{\la\o 4!}\[\(\vf +v,\vf +v\)-(v,v)\]^2\]
\ ,\label{sact}\ee
is a natural  term of the first approximation gauge invariant vertex functional;
 notice that the coefficient $\la$ is dimensionless and:
\be {\d\C_s\o\d\vf}|_{\vf=0}=0\ ,\label{vac}\ee
as it is required by (\ref{10}) and (\ref{pm})

If the $\t$ representation of the gauge group is reducible, every irreducible
component of it carries an invariant scalar product and hence the second term
of $\C_s$ is replaced by a polynomial of second degree in these scalar products,
 bounded from above and satisfying (\ref{vac}).

Furthermore, consider the commutator:
\begin{eqnarray} \left[ D_\mu ,D_\nu \right]\( \vf +v\)
 & = & \left[ \partial_\mu - i A_\mu^{\alpha}
\tau_\alpha ,\partial_\nu - i A_\nu^{\beta}
\tau_\beta \right]\( \vf +v\) \nonumber
\\
& = & -i \tau_\alpha \left[ \partial_\mu A_\nu^{\alpha} -
\partial_\nu A_\mu^{\alpha}
+ A_\mu^{\beta} A_\nu^{\gamma} f_{\beta \gamma}^{\alpha} \right]\( \vf +v\)
 \nonumber \\
& \equiv & i \tau_\alpha G_{\mu \nu}^{\alpha}\( \vf +v\)
\equiv G_{\mu \nu}\( \vf +v\)\ .\label{fst} \end{eqnarray}
$ G_{\mu \nu}$ is the "field strength" of our gauge theory.

It is clear that:
\be \delta G_{\mu \nu} = i \left[ \Lambda^\beta \tau_\beta,
G_{\mu \nu} \right]\ ,\label{tfst} \ee
since, according to (\ref{cd}),
 an analogous equation holds true for any product of covariant derivatives;
thus the terms of the form:
\be Tr\(\[D_{\n_1},..,\[D_{\n_{k_1}},G_{\mu_1\nu_1}\]..\]..
\[D_{\n_j},..,\[D_{\n_{k_l}},G_{\mu_j\nu_j}\]..\]\)\ ,\label{trac}\ee
are gauge invariant.
In particular the functional:
\be\C_g\equiv - k\ \inx Tr\(G_{\mu \nu}G^{\mu \nu}\)\ ,\label{gact}\ee
is the natural candidate for a first approximation to the gauge field vertex
functional of the effective theory.
If the compact gauge group is the direct product of a set of invariant
factors (either abelian or simple), the field strength decomposes into the
sum of the contributions of the single factors:
\be G_{\mu \nu}=\sum_IG_{\mu \nu}^{(I)}\ ,\ee
and (\ref{gact}) is replaced by the sum of the contributions of the factors,
each multiplied by an independent coefficient $k_I$.

The sum of  (\ref{gact}) and  (\ref{sact}) gives the complete effective field
approximation vertex functional. To have a better understanding
of the physical content of our theory we now discuss in some detail an
 example in which the gauge group is $SU(2)$ and the scalar representation
is the fundamental one:
\be \vf = \left( \begin{array}{c} \vf_{+} \\ \vf_{-} \end{array} \right) =
1/\sqrt{2} \left( \begin{array}{cc} \vf_{2}+i\vf_{1} \\ \vf_{4}+i\vf_{3}
\end{array} \right)\ . \label{scal}\ee
The field strength is:
\be G_{\mu \nu}^{i} = \partial_\mu A_\nu^{i} - \partial_\nu A_\mu^{i} +
\sqrt{2}
\epsilon_{ijk} A_\mu^{j}A_\nu^{k}\ .\label{fst1}\ee
Choosing the constant $k$ in (\ref{gact}) according:
\be k={1\o4g^2}\ ,\label{coupl}\ee
and substituting the isovector potential:
\be {\vec A}_\mu\rightarrow g\ \ca_\mu\ ,\label{sub}\ee (we use arrows to
single out isovectors) we have, in the inhomogeneous case, the effective vertex
functional: \bea\C =  \inx\[\left( \partial_\mu \vf^{\dagger}
- i \(\vf+v\)^{\dagger} g  { \ca_\mu} \cdot \frac{ {\vec\sigma}}{2},
\partial^\mu \vf + i g  { \ca^\mu} \cdot \frac{ {\vec\sigma}}{2}
\(\vf+v\) \right)  \acca -{\la\o 4!}\[\(\(\vf +v\)^{\dagger},\vf +v\)-
(v^{\dagger},v)\]^2
-{1\o4} \left(\partial_\mu  \ca_\nu
- \partial_\nu  \ca_\mu \right)\cdot\(\partial^\mu  \ca^\nu
- \partial^\nu  \ca^\mu\) \acca
- g \sqrt{2} \left(\partial_\mu  \ca_\nu\cdot  \ca_\mu\wedge  \ca_\nu
\right)  - {g^2\o2}\( \ca_\mu^{2}  \ca_{\nu}^{2} - \left(
 \ca_\mu \cdot \ca_\nu \right)^{2}\)  \]\ .\label{av}\eea
It is interesting to compare this functional
with the effective vertex functional in the case $v= 0$.
Indeed in this situation one has:
\bea\C =  \inx\[\left( \partial_\mu \vf^{\dagger}
- i \vf^{\dagger} g  { \ca_\mu} \cdot \frac{ {\vec\sigma}}{2},
\partial^\mu \vf + i g  { \ca^\mu} \cdot \frac{ {\vec\sigma}}{2}
\vf \right)  \acca -{m^2\o2}  \left( \vf^{\dagger}\vf \right)^2
-{\la\o4!}\left( \vf^{\dagger}\vf \right)^4
-{1\o4} \left(\partial_\mu  \ca_\nu
- \partial_\nu  \ca_\mu \right)\cdot\(\partial^\mu  \ca^\nu
- \partial^\nu  \ca^\mu\) \acca
- g \sqrt{2} \left(\partial_\mu  \ca_\nu\cdot  \ca_\mu\wedge  \ca_\nu
\right)  - {g^2\o2}\( \ca_\mu^{2}  \ca_{\nu}^{2} - \left(
 \ca_\mu \cdot \ca_\nu \right)^{2}\)  \]\ .\label{a0}\eea
Notice that the second term in the right-hand side of (\ref{av})
has been replaced by two independent term in (\ref{a0}) containing the new
parameter $m^2$; as a matter of fact this new parameter compensates the loss
of $v$. Owing to gauge invariance that makes equivalent all the choices
 of $v$ with the same norm, $v$ is equivalent to a single parameter.

To study the particle content of these theories we select the two-point
vertex generators. In the inhomogeneous case we choose a particular $v$
according:
\be v = \left( \begin{array}{c} 0 \\ F/\sqrt{2}\end{array} \right)\ ,\label{vch}
 \ee
with $F$ real and positive. Introducing the symbol $\vec \vf$ for $\(\vf_1,\vf_2
, \vf_3\)$ we get:
\bea\Gamma_2 =\int d^4x\left[\frac{1}{2}  \(\vec{{\cal A}}_{\mu} (g^{\mu \nu}
\partial^2 -
{\partial}^{\mu} {\partial}^{\nu} )\cdot\vec{{\cal A}}_{\nu}\) +
\acca
{F\over2}g \(\vec{{\cal A}}_{\mu}\partial^{\mu}\cdot\vec\varphi\) +
{(\partial {\varphi})^2\over 2} +
{F^2g^2{\cal A}^2\over8}+
{(\partial \varphi_4)^2\over 2} -\lambda F^2\varphi^2_4\right]\ ,
\label{hi}\eea
that can be simplified by the substitution:
\be \ca + {2\o gF}\p \vec\vf\rightarrow\ca\ ,\label{subs}\ee
leading to:
\bea\Gamma_2 = \int d^4x\left[-\frac{1}{2} \( \vec A_{\mu} (g^{\mu \nu}
\partial^2 -
{\partial}^{\mu} {\partial}^{\nu} )\cdot\vec A_{\nu}\) +\acca.
{F^2g^2A^2\over8}+
{(\partial \varphi_4)^2\over 2} -\lambda F^2\varphi^2_4\right]\
.\label{hii}\end{eqnarray}
In the homogeneous case we have directly:
\be
\Gamma_2=\int d^4 x \frac{1}{2} \left [- \(\vec{{\cal A}}_{\mu} (g^{\mu \nu}
\partial^2 -
{\partial}^{\mu} {\partial}^{\nu} )\cdot\vec{{\cal A}}_{\nu}\) +
(\partial \vec{\varphi})^2 +
(\partial \varphi_4)^2-
m^2 ( \vec{\varphi}^2+\varphi_4^2 ) \right ] \ .\label{go}
\ee
To both functionals we add the auxiliary (\ref{xi}) term introducing
for simplicity the Landau gauge choice:
$\xi\rightarrow\infty$ that corresponds to the constraint:
\be\p^\mu\ca_\mu =0\ .\label{land}\ee
It is clear that (\ref{hii}) describes a free isovector-vector field
with mass $m=\sqrt{2}\la F$ and a neutral scalar field with mass $M={gF\o2}$.
Indeed the vector field is quantized by a decomposition strictly analogous to
that in (\ref{asy}) with the choice (\ref{tr}) of the polarization vectors.

 In the homogeneous case we introduce the decomposition:
\begin{equation}
\ca^{\mu}(x)= \frac{1}{(2\pi)^{3/2}}\sum_{\lambda}
\int\frac{d^3p}{\sqrt{2p}}\left( \epsilon^{\mu}_{\lambda,\vec{p}}
\ca_{\lambda\vec{p}} e^{-ipx}+hermitian \ \  conjugate\right)\ ,
\end{equation}
where $\epsilon^{\mu}_{\lambda,\vec{p}}$ give a four-vector basis orthogonal
to $p_\mu$, that is:
\begin{eqnarray}
\epsilon_{3,\vec{p}}^{\mu}& =& \frac{ p^{\mu}}{p_0} \nonumber \\
\epsilon_{i,\vec{p}}^{\mu}\ \  &\mbox{i=1,2}&
\left\{ \begin{array}{rcl}
 (\epsilon_i,\epsilon_j) & = & -\delta_{ij} \\
 (\epsilon_i,p)    & = & (\epsilon_i,\bar p) =   0  \\
\end{array} \right.
\end{eqnarray}
where $\bar p$ is the image of $p$ after a parity reflection.
The two polarization vectors $\epsilon_i$ for $i=1,2$ correspond
 to isovector states with helicity $\la=\pm 1$. On the contrary
 $\epsilon_3$ does not
correspond to any dynamical degree of freedom, indeed, owing to gauge
invariance,
the corresponding term of the vector field does not contribute to the action.
Thus the third polarisation corresponds to the freedom of redefining the vector
field by a gauge transformation. Notice that in the massless case gauge
invariance is needed to guarantee the decoupling of the states with this
polarization. Furthermore the functional (\ref{go}) describes four real scalar
fields with mass $m$.

 Therefore, comparing the helicity states
of the two theories, we find two isovector states with non-vanishing helicities
in  both cases. They are however massless in the homogeneous case and massive
in the inhomogeneous one. There are furthermore four massive, helicity zero,
states in both theories.

Notice that, from the point of view of symmetry, both theories have gauge
 invariant quantum actions, however in the inhomogeneous case the vacuum
state, that corresponds to the vanishing field configuration, is not gauge
invariant. In this situation one speaks of "spontaneous symmetry breakdown". In
conclusion, the spontaneous breakdown of gauge invariance has made the vector
field massive thus transferring an isotriplet of helicity zero states from the
scalar to the vector fields. This is the "Higgs mechanism".

It is quite clear from the above construction that, if the gauge symmetry
remains
 unbroken, the vector fields are massless.

To conclude our analysis of an effective vector field, let us notice that the
use we have made of an effective theory in the massless case is questionable;
indeed as we have said above, the use of an effective approximation to describe
the vertex functional is justified by analyticity arguments at vanishing
momenta that do not hold in the massless case. Indeed in this case one has to
take into account the singularities corresponding to the intermediate states of
massless particles. In the confined theory it is believed that these "infrared"
singularities wash out the vector asymptotic states. \salto
\sec{The functional integral construction}
\salto
The functional integral construction of  field theories is based on the
identification of suitable measures in the functional space of fields analogous
to the heuristic form (\ref{er}). At first sight the choice of this measure
should be completely determined on the basis of equation (\ref{16}) that
implies that the classical action is identified with the classical
approximation to the vertex functional; therefore the quantization criterion
should be the gauge invariance of $S$. However this possibility is frustrated
by the fact that if the measure is gauge invariant the corresponding Feynman
integral does not exist. In perturbation theory this is due to the lack of
uniqueness of the propagator. Considering as reference example a SU(2) pure
Yang-Mills theory, whose action corresponds to (\ref{hii}) deprived of the
scalars, we have the Lagrangian density:
\be
{\cal L}=-\frac{1}{4} \vec{G}_{\mu \nu} \vec{G}^{\mu \nu}
\ ,\label{lag}\ee
and the propagator should be defined by the equation:
\be
(g^{\mu \rho} \partial^2 - \partial^{\mu} \partial^{\rho})\D_{\rho\nu}(x)
=-\d^\mu_{\nu}\d (x)\ , \label{propag}
\ee
it is however not unique, since for any function $F(x)$:
$(g^{\mu \rho} \partial^2 - \partial^{\mu} \partial^{\rho})\p_{\rho}F(x)=0$.

It is
perhaps worth reminding here that the functional integral construction
of $Z$ is based on the so-called Wick rotation. The theory is transformed to
imaginary time; correspondingly the measure $e^{{i\o\hbar}S}$ turns into
$e^{-{S_e\o\hbar}}$, where $S_e$ is bounded from below and it is Euclidean
rather than Lorentz invariant. In this Euclidean formulation the Feynman
integral can be defined if $S_e$ increases rapidly enough in
all the directions of the field functional space. The identification of
the measure should lead to a well defined functional generator of the
Euclidean Green functions that turn out to be the analytic continuation of the
physical (Minkowskian) ones. Therefore, in reality, when one speaks of the
Feynman functional integral, one has in mind the generator of the Euclidean
theory. From the perturbative point of view the whole procedure of Wick
rotation combined with analytic continuation is in a sense trivial since the
Feynman amplitudes are explicit analytic functions of the momenta.

Coming back to our theory, that we convert into its Euclidean version, gauge
invariance makes the measure constant along the paths in the field functional
space spanned by the transformations (\ref{ino}) that are called "orbit" of a
given configuration. The lack of convergence in the orbit directions makes the
functional integral ill-defined.

One could suggest here to provide the wanted convergence introducing into $S$
the $\xi$ term (\ref{xi}) corresponding in perturbation theory to a particular
choice of the propagator, that is:
\be \Delta^{\mu \nu}=\frac{1}{(2 \pi)^4} \int dk
\frac{e^{\imath k x}}{k^2+\imath 0^{+}}(g^{\mu
\nu}+\(\xi^{-1}-1\)\frac{k^{\mu} k^{\nu}} {k^2})\ .\label{propxi}\ee
 As a matter of fact, this is only correct in
the abelian case (QED). In the general case the introduction of a $\xi$ term
 is allowed into $\C$ since, as we have shown above, the $\S$ matrix remains
$\xi$-independent. However this argument cannot be automatically extended to the
case in which one inserts the $\xi$ term into the action
$S$. Indeed the proof of the $\xi$-independence depends crucially on the
Legendre transformation connecting $Z_c$ to $\C$. This Legendre transformation
can be interpreted as a development in Feynman tree diagrams, that is in
diagrams without loops. On the contrary $Z_c$ is related to $S$ by a complete
development into Feynman diagrams among which those with loops violate the
$\xi$-independence.

 The correct version of the $\xi$ term has been discovered
by Faddeev and Popov \cite{fp}.
 We are now going to describe their result; however, instead of the $\xi$ term,
we shall discuss the introduction of the term \be\inx{\vec\la}\p^\mu\ca_\mu\
,\label{landa}\ee that corresponds to it in the limit: $\xi\rightarrow\infty$.
In (\ref{landa}) there appears the auxiliary isovector-scalar field $\vec\la$
that is called the Lautrup-Nakanishi field.  Considering the variation of this
term under an infinitesimal gauge transformation we get:
\be\inx \lambda \partial^2 {\Lambda}\ ,\label{gtt1}\ee
in the abelian case, where we have forgotten the arrows, and:
\be\inx
\vec\lambda \left [\partial^2 \vec{\Lambda}
-g \sqrt{2}\partial_{\mu}(\vec{{\cal A}}^{\mu} \wedge \vec{\Lambda})
\right]\equiv\inx
\vec\lambda \partial D \vec{\Lambda}
 \ ,\label{gtt2} \ee
in the non-abelian one.

Let us for a moment interpret the gauge parameter $\Lambda$ as a field
component. We see from (\ref{gtt1}) that in the abelian case it is a free field
conjugate to $\la$ and hence it is not involved into the dynamics. On the
contrary, in the non-abelian case (\ref{gtt2}), $\Lambda$ is coupled to the
gauge field; therefore the introduction of the new term (\ref{landa}) modifies
the dynamics in a substantial way thus violating the decoupling theorem (the
$\xi\rightarrow\infty$ limit of the $\xi$ independence).

Naively we can say that Faddeev-Popov have cured this sickness compensating
the two fake degrees of freedom ($\vec\la$ and $\vec\Lambda$) by the
introduction of their compensating antiquantized images.

In the case of ordinary integrals one can understand
this compensation in the following way: consider a (finite) system of
Grassmann (anticommuting) variables $\z_i$ and $\bar\z_j$:\be
\left \{
\begin{array}{l}
\zeta_i \bar{\zeta}_j+\bar{\zeta}_j\zeta_i=0 \\
\zeta_i \zeta_j+\zeta_j\zeta_i=0 \\
\bar{\zeta}_i \bar{\zeta}_j+\bar{\zeta}_j\bar{\zeta}_i=0
\end{array}
\right.\label{grass}
\ee
and notice that a generic function of then is linear in each variable, that is,
 concerning e.g. $\z_k$, a generic function can be written in the form:
\be
F(\zeta,\bar{\zeta})=A_k(\zeta,\bar{\zeta})+\zeta_k B_k(\zeta,\bar{\zeta})\
.\label{funct}\ee We define the partial derivative:
\be \partial_{\z_k}F(\zeta,\bar{\zeta})
\equiv  B_k(\zeta,\bar{\zeta})\ .\label{partial}\ee
$\p_{\bar {\z}_i}$ is defined analogously. It is clear that:
\bea \p_{\z_i}\z_j+\z_j\p_{\z_i}=\d_{ij}\ ,\nn
\p_{\z_i}{\bar\z}_j+{\bar\z}_j\p_{\z_i}=0\ ,\label{anticom}\eea
The set of these partial derivatives generates a new Grassmann algebra
analogous and conjugate to (\ref{grass}). Owing to the fact that $\p_{\z_k}F$
is independent of $\z_k$ it is possible, and perhaps natural, to define the
Berezin integral:
\be
\int d\zeta_k F(\zeta,\bar{\zeta})\equiv
\sqrt{\pi}^{-1} {\partial\over\partial \zeta_k}F(\zeta,\bar{\zeta})
= \sqrt{\pi}^{-1} B_k(\zeta,\bar{\zeta}) \ .\label{berez}\ee
Now, given a generic matrix $M$ we can compute the Berezin integral:
\bea\int  d\z d{\bar\z}e^{\bar{\zeta}_iM_{ij}\zeta_j}\equiv
\int d\zeta_1 \ldots d\zeta_n d \bar{\zeta}_n \ldots d \bar{\zeta}_1
e^{\bar{\zeta}_iM_{ij}\zeta_j}\nn
=\int d\zeta_1 \ldots d\zeta_n d \bar{\zeta}_n
\ldots d \bar{\zeta}_1  \frac{(\bar{\zeta} M \zeta)^n}{n!}=
{\det M\o\pi^n} \label{ib}
\ .\eea
If $M$ is positive (it is diagonalizable with positive eigenvalues) we have
also:
\be
\int \prod_i d z_i d \bar{z}_i
e^{-\bar{z}_i M_{ij} z_j}=\frac{\pi^n}{\det M}\ ,
\label{gauss}\ee
therefore, combining (\ref{ib}) and (\ref{gauss}) we have:
\be
\int \prod_{i=1}^n d z_i d \bar{z}_j e^{-\bar{z}_l M_{lm} z_m}
\cdot \int \prod_{j=1}^n d \zeta_j d \bar{\zeta}_j
e^{-\bar{\zeta}_k M_{kn} \zeta_n}=
1\ .
\label{uno}\ee
This equation is naturally extended to the functional integrals:
 considering in particular the bosonic field (ordinary functional variables)
$\vec\la$ and $\vec \Lambda$ and the corresponding fermionic images
(Grassmannian functional variables) $\vec{\bar\u}$ and $\vec\u$ one has:
\be
\int \prod_{x} d \vec {\lambda}(x) d \vec{\Lambda}(x)
d \vec{\omega}(x) d \vec{\bar{\omega}}(x)
e^{\int dx [\lambda \partial_{\mu} D^{\mu}\Lambda + \imath \bar{\omega}
\partial_{\mu}D^{\mu}\omega}]=1\ . \label{compen}
\ee
We have assumed that the operator $\p D$ be negative definite. As a matter of
fact Gribov \cite{3} has shown that this is not true for all choices of vector
fields; it is certainly true for small fields and hence our formula is correct
at least in perturbation theory.

The Faddeev-Popov trick consists in the choice of the action:
\be S_e=\inx  \[\frac{1}{4} \vec{G}_{\mu \nu}\cdot\vec{G}^{\mu \nu}
-i\vec{\lambda}\cdot\partial_{\mu}\vec{{\cal A}}^{\mu}-
 \vec{\bar{\omega}}\cdot
\partial_{\mu}D^{\mu}\vec{\omega}\]\ ,\label{fpact}\ee
and hence in that of the Green functional:
 \be
Z=\int d\mu_C\
e^{i\int dx [
\vec{\lambda}\cdot\partial_{\mu}\vec{{\cal A}}^{\mu}-i
 \vec{\bar{\omega}}\cdot
\partial_{\mu}D^{\mu}\vec{\omega}]}
e^{i\int dx [{\vec j}^\mu\cdot\ca_\mu +{\vec
J}\cdot{\vec\la} +{\vec
\eta}\cdot{\vec\u}+{\vec{\bar\eta}
}\cdot{\vec{\bar\u}}]}\ , \label{vacfun}\ee where we have defined the
functional measure: \be d\mu_C\equiv N \prod_{x} d \vec {\lambda}(x) d\ca(x)
d \vec{\omega}(x) d \vec{\bar{\omega}}(x)
e^{\int dx \[-\frac{1}{4} \vec{G}_{\mu \nu}\cdot\vec{G}^{\mu \nu}\]}\
.\label{measc}\ee
According to (\ref{compen}) the introduction of two antiquantized isovector
fields $\vec{\bar\u}$ and $\vec\u$ exactly compensates in the functional
integral the contribution of the gauge degrees of freedom of the vector fields
and of the Nakanishi-Lautrup field. In a more conventional language one says
that the contributions to the transition probabilities of the states of pairs
$\vec{\bar\u}$-$\vec\u$, that have negative norm, exactly compensate those of
pairs of fake particles corresponding to $\vec\la$ and $\vec \Lambda$.

From a mathematical point of view this consists in the
insertion into the functional measure of an invariant $\d$ function
implementing the gauge condition $\p\ca=0$ \cite{bi}.
\salto\sec{Symmetry
properties of the Green functional} \salto
In order the Faddeev-Popov trick to work correctly it is crucial that the
third term in the action (\ref{fpact}) correspond exactly to the gauge variation
of the second one. To control this correspondence is a difficult task since the
action has to be modified by the regularization and renormalization procedure
that we do not discuss in these lectures. It is therefore essential to have a
renormalization criterion that guarantees the persistence of the above
correspondence.  This criterion is identified with the invariance condition of
the action under a non-linear system of transformations (BRST) \cite{brs}
that, for the above action, are written: \bea
\vec{{\cal A}}_{\mu}  \rightarrow \vec{{\cal A}}_{\mu}-\varepsilon
D_{\mu}\vec{\omega}\equiv\ca_\mu-\varepsilon s \ca_\mu\nn
\vec{\bar{\omega}}  \rightarrow \vec{\bar{\omega}}-i\varepsilon \vec{\lambda}
\equiv{\vec{\bar\u}}-\varepsilon s {\vec{\bar\u}}\nn
\vec{\omega}  \rightarrow \vec{\omega}-g
\frac{\varepsilon}{\sqrt{2}}\vec{\omega} \wedge \vec{\omega}
\equiv\vec\u-\varepsilon s\vec\u\ ,\label{brs}
\eea
where $\varepsilon$ is a Grassmannian parameter.
It is easy to verify that the
$s$ operator defined in (\ref{brs}) together with the further condition:
$s\vec\la=0$ is nilpotent: \be s^2\equiv 0\ .\label{nilpo}\ee Indeed, for
example: \be s\vec\u\propto
s(\vec{\omega} \wedge \vec{\omega})\propto
(\vec{\omega} \wedge \vec{\omega})\wedge\vec{\omega}=0 \ ,
\ee
The invariance of the action is verified considering that the $s$ operator
acts on the vector field in much the same way as an infinitesimal gauge
transformation and hence:
\be s  \vec{G}_{\mu \nu}\cdot\vec{G}^{\mu \nu} =0\ .\ee
Furthermore we have:
\be
s \vec{\lambda}\cdot\partial_{\mu}\vec{{\cal A}}^{\mu} ={\varepsilon} \vec\la
\cdot\partial_{\mu}  D^{\mu} \vec{\omega}\ ,\ee and on account of the nilpotency
of $s$:
\be s \vec{\bar{\omega}}\cdot\partial_{\mu}D^{\mu}\vec{\omega}=
-i \varepsilon
\vec{\lambda}\cdot \partial_{\mu} D^{\mu}\vec{\omega}\ .\ee
One has to keep in mind that nilpotency is an essential condition to extend
BRST symmetry at the renormalized level.

 In terms of the $s$ operator the
Faddeev-Popov insertion (the second and third terms in (\ref{fpact})) is written
$s\inx\({\vec{\bar\u}}\cdot\p\ca\)$. The invariance of this expression is an
obvious consequence of the nipotency of $s$. The
Faddeev-Popov insertion can be generalized reintroducing
the $\xi$ parameter according:\be s\inx
 {\vec{\bar\u}}\cdot\(\p\ca +{i\o2\xi}\vec\la\)\ .\label{neoxi}\ee
Indeed the introduction of the new term corresponds to the substitution:
\be S_e\rightarrow S_e +{1\o2\xi}\inx\ \la^2 \ .\label{xilag}\ee
Separating a quadratic term in $\vec\la+i\xi\p\ca$, this action corresponds to
the Lagrangian density:
\be{\cal L}=\frac{1}{4} \vec{G}_{\mu \nu}\vec{G}^{\mu \nu}+{\xi\over 2}
\partial \vec{{\cal A}} \cdot \partial \vec{{\cal A}}-
\partial_{\mu}\vec{\bar{\omega}}D^{\mu}\vec{\omega}\ .\label{lagran}\ee
Let us assume that the functional measure (\ref{measc}) be $s$-invariant;
this, of course, depends on the regularization-renormalization procedure.
Given any integrable functional $\Xi$ one has the  Slavnov-Taylor identity
\cite{st}: \be \int d\mu_C e^{-s\inx\
 {\vec{\bar\u}}\cdot\(\p\ca +{i\o2\xi}\vec\la\)} s\Xi=0\ ,\label{sti}\ee
indeed:
\be \int d\mu_C e^{-s\inx\
 {\vec{\bar\u}}\cdot\(\p\ca +{i\o2\xi}\vec\la\)} s\Xi=\int d\mu_C s\[
e^{-s\inx\
 {\vec{\bar\u}}\cdot\(\p\ca +{i\o2\xi}\vec\la\)} \Xi\]=0\ ,\ee
the last identity following from the invariance of the measure.

The meaning of this identity is the vanishing of all the  correlators between
$s$-invariant operators and elements of the image of $s$, that appear above as
$s\Xi$. Indeed, if $s\Sigma=0$, one has:
\be\int d\mu_C e^{-s\inx\
 {\vec{\bar\u}}\cdot\(\p\ca +{i\o2\xi}\vec\la\)}\Sigma s\Xi=
\int d\mu_C e^{-s\inx\
 {\vec{\bar\u}}\cdot\(\p\ca +{i\o2\xi}\vec\la\)} s\(\Sigma\Xi\)=0\ .\ee
Now it is not difficult to understand how BRST symmetry is related to
$\xi$-independence.
The observables of gauge theories correspond to gauge and hence
$s$ invariant operators; that is: if the operator $\Omega$ is observable and
hence physically meaningful, then: $s\Omega=0$. As a matter of fact
 $s$-invariance,
modulo the image of $s$, gives a generalized definition of observables in gauge
theories.

 Using (\ref{sti}) we can
prove the $\xi$-independence  of the  vacuum expectation value of $\Omega$.
Indeed computing the $\xi$-derivative:
\bea\p_\xi \int d\mu_C e^{-s\inx\
 {\vec{\bar\u}}\cdot\(\p\ca +{i\o2\xi}\vec\la\)}\Omega\nn
={i\o2\xi^2} \int d\mu_C e^{-s\inx\
 {\vec{\bar\u}}\cdot\(\p\ca
+{i\o2\xi}\vec\la\)}\Omega s\inx{\vec{\bar\u}}\cdot\vec\la\nn=
{i\o2\xi^2} \int d\mu_C e^{-s\inx\
 {\vec{\bar\u}}\cdot\(\p\ca
+{i\o2\xi}\vec\la\)}s\(\Omega \inx{\vec{\bar\u}}\cdot\vec\la\)=0\ .\eea
The last identity following from the Slavnov-Taylor identity.

This result generalizes that of the $\xi$-independence of the
$\S$ matrix since, modulo infrared effects, the sources of the asymptotic field
of the vector  and matter particles, that we have written in the form:
$K\f_{in}$, are observable operators.
Using the Slavnov-Taylor identity it is also possible to give a direct proof of
the $\S$-matrix unitarity \cite{bi}.

\salto
\salto


\begin{thebibliography}
{15}

\bibitem{bu}

D. Buchholz,  Nucl.Phys. B469 (1996) 333.

\bibitem{ym}

C.N. Yang and R.L. Mills, Phys. Rev. 95 (1954) 631.

\bibitem{2}

K.G.Wilson, Phys.Rev. D10 (1974) 2445.



\bibitem{1}
C.Itzykson e J.-B.Zuber, Quantum Field Theory, MacGraw-Hill, New-York 1980.

\bibitem{be}
see e.g.:

     C.Becchi, Lectures on the renormalization of gauge theories, in:
     Relativity, Groups and Topology II (les Houches 1983) B.S. DeWitt
     and R.Stora Eds. (Elsevier Science Pub.B.V. 1984).

\bibitem{ls}

C.H. Llewellyn Smith, Phys. Letters 46B (1973) 233.


\bibitem{3}
V. N. Gribov, ``Instability of non-abelian gauge theories and
impossibility of choice of Coulomb gauge'', SLAC Translation 176, (1977).

I.M. Singer
Commun.Math.Phys. 60 (1978) 7.

\bibitem{ls}

C.H. Llewellyn Smith, Phys. Letters 46B (1973) 233.

\bibitem{ds}

G. Parisi, Phys. Rev. D11 (1975) 970.

G. t'Hooft, Proceedings of the EPS International Conference, Palermo (1975).

S. Mandelstam, Phys. Reports 23 (1976) 245.

\bibitem{fp}
L.D.Faddeev and V.N.Popov, Phys.Letters B25 (1967) 29.

\bibitem{bi}

 C. Becchi,
Introduction to BRS symmetry; hep-th/9607181 23 Jul 1996.


\bibitem{brs}
C.Becchi, A.Rouet and R.Stora,  Phys Letters. B32 (1974) 344.

C.Becchi, A.Rouet and R.Stora,  Commun.Math. Phys. 42 (1975) 127.

L.V.Tyutin, Lebedev preprint FIAN n.39 (1975).

C.Becchi, A.Rouet and R.Stora,  Ann. Phys. 98 (1976) 287

\bibitem{st}
A.A.Slavnov, Theor. Math. Phys. 10 (1972) 99.

J.C.Taylor, Nucl.Phys. B33 (1971) 436.


\end{thebibliography}
\end{document}